\documentclass[twocolumn,showpacs,preprintnumbers,amsmath,amssymb,superscriptaddress,amssymb]{revtex4-2}
\usepackage{subcaption}
\usepackage{graphicx}
\usepackage{epsfig}
\usepackage{multirow}
\usepackage{pbox}
\usepackage{array}
\usepackage{relsize}
\usepackage{siunitx}
\usepackage{braket}
\usepackage[labelfont=normalfont,textfont=normalfont,singlelinecheck=off,justification=justified]{caption}

\begin{document}
\newcolumntype{P}[1]{>{\centering\arraybackslash}p{#1}}

\title{Study of one-proton transfer reaction for the $^{18}$O + $^{48}$Ti system at 275 MeV}

\author{O. Sgouros}
\email{onoufrios.sgouros@lns.infn.it}
\affiliation{INFN - Laboratori Nazionali del Sud, Catania, Italy}
\author{M. Cavallaro}
\affiliation{INFN - Laboratori Nazionali del Sud, Catania, Italy}
\author{F. Cappuzzello}
\affiliation{INFN - Laboratori Nazionali del Sud, Catania, Italy}
\affiliation{Dipartimento di Fisica e Astronomia "Ettore Majorana", Universit\`a di Catania, Catania, Italy}
\author{D. Carbone}
\affiliation{INFN - Laboratori Nazionali del Sud, Catania, Italy}
\author{C. Agodi}
\affiliation{INFN - Laboratori Nazionali del Sud, Catania, Italy}
\author{A. Gargano}
\affiliation{INFN - Sezione di Napoli, Napoli, Italy}
\author{G. De Gregorio}
\affiliation{INFN - Sezione di Napoli, Napoli, Italy}
\affiliation{Dipartimento di Matematica e Fisica, Universit\`a della Campania "Luigi Vanvitelli", Caserta, Italy}
\author{C. Altana}
\affiliation{INFN - Laboratori Nazionali del Sud, Catania, Italy}
\author{G. A. Brischetto}
\affiliation{INFN - Laboratori Nazionali del Sud, Catania, Italy}
\affiliation{Dipartimento di Fisica e Astronomia "Ettore Majorana", Universit\`a di Catania, Catania, Italy}
\author{S. Burrello}
\affiliation{Universit\'e Paris-Saclay, CNRS/IN2P3, IJCLab, Orsay, France}
\affiliation{Technische Universit{\"a}t Darmstadt, Institut f{\"u}r Kernphysik, Darmstadt, Germany}
\author{S. Calabrese}
\affiliation{INFN - Laboratori Nazionali del Sud, Catania, Italy}
\affiliation{Dipartimento di Fisica e Astronomia "Ettore Majorana", Universit\`a di Catania, Catania, Italy}
\author{D. Calvo}
\affiliation{INFN - Sezione di Torino, Torino, Italy}
\author{V. Capirossi}
\affiliation{INFN - Sezione di Torino, Torino, Italy}
\affiliation{DISAT - Politecnico di Torino, Torino, Italy}
\author{E. R. Ch\'avez Lomel\'i}
\affiliation{Instituto de F\'isica, Universidad Nacional Aut\'onoma de M\'exico, Mexico City, Mexico}
\author{I. Ciraldo}
\affiliation{INFN - Laboratori Nazionali del Sud, Catania, Italy}
\affiliation{Dipartimento di Fisica e Astronomia "Ettore Majorana", Universit\`a di Catania, Catania, Italy}
\author{M. Cutuli}
\affiliation{INFN - Laboratori Nazionali del Sud, Catania, Italy}
\affiliation{Dipartimento di Fisica e Astronomia "Ettore Majorana", Universit\`a di Catania, Catania, Italy}
\author{F. Delaunay}
\affiliation{INFN - Laboratori Nazionali del Sud, Catania, Italy}
\affiliation{Dipartimento di Fisica e Astronomia "Ettore Majorana", Universit\`a di Catania, Catania, Italy}
\affiliation{LPC Caen, Normandie Universit\'e, ENSICAEN, UNICAEN, CNRS/IN2P3, Caen, France}
\author{H. Djapo}
\affiliation{Ancara University, Institute of Accelerator Technologies, Turkey}
\author{C. Eke}
\affiliation{Department of Mathematics and Science Education, Faculty of Education, Akdeniz University, Antalya, Turkey}
\author{P. Finocchiaro}
\affiliation{INFN - Laboratori Nazionali del Sud, Catania, Italy}
\author{M. Fisichella}
\affiliation{INFN - Laboratori Nazionali del Sud, Catania, Italy}
\author{A. Foti}
\affiliation{INFN - Sezione di Catania, Catania, Italy}
\author{A. Hacisalihoglu}
\affiliation{Institute of Natural Sciences, Karadeniz Teknik Universitesi, Trabzon, Turkey}
\author{F. Iazzi}
\affiliation{INFN - Sezione di Torino, Torino, Italy}
\affiliation{DISAT - Politecnico di Torino, Torino, Italy}
\author{L. La Fauci}
\affiliation{INFN - Laboratori Nazionali del Sud, Catania, Italy}
\affiliation{Dipartimento di Fisica e Astronomia "Ettore Majorana", Universit\`a di Catania, Catania, Italy}
\author{R. Linares}
\affiliation{Instituto de F\'isica, Universidade Federal Fluminense, Niter\'oi, Brazil}
\author{J. Lubian}
\affiliation{Instituto de F\'isica, Universidade Federal Fluminense, Niter\'oi, Brazil}
\author{N. H. Medina}
\affiliation{Instituto de F\'isica, Universidade de S\~ao Paulo, S\~ao Paulo, Brazil}
\author{M. Moralles}
\affiliation{Instituto de Pesquisas Energeticas e Nucleares IPEN/CNEN, S\~ao Paulo, Brazil}
\author{J. R. B. Oliveira}
\affiliation{Instituto de F\'isica, Universidade de S\~ao Paulo, S\~ao Paulo, Brazil}
\author{A. Pakou}
\affiliation{Department of Physics, University of Ioannina and Hellenic Institute of Nuclear Physics, Ioannina, Greece}
\author{L. Pandola}
\affiliation{INFN - Laboratori Nazionali del Sud, Catania, Italy}
\author{F. Pinna}
\affiliation{INFN - Sezione di Torino, Torino, Italy}
\affiliation{DISAT - Politecnico di Torino, Torino, Italy}
\author{G. Russo}
\affiliation{Dipartimento di Fisica e Astronomia "Ettore Majorana", Universit\`a di Catania, Catania, Italy}
\affiliation{INFN - Sezione di Catania, Catania, Italy}
\author{M. A. Guazzelli}
\affiliation{Centro Universitario FEI, S\~ao Bernardo do Campo, Brazil}
\author{V. Soukeras}
\affiliation{INFN - Laboratori Nazionali del Sud, Catania, Italy}
\author{G. Souliotis}
\affiliation{Department of Chemistry, University of Athens and Hellenic Institute of Nuclear Physics, Athens, Greece}
\author{A. Spatafora}
\affiliation{INFN - Laboratori Nazionali del Sud, Catania, Italy}
\affiliation{Dipartimento di Fisica e Astronomia "Ettore Majorana", Universit\`a di Catania, Catania, Italy}
\author{D. Torresi}
\affiliation{INFN - Laboratori Nazionali del Sud, Catania, Italy}
\author{A. Yildirim}
\affiliation{Department of Physics, Akdeniz Universitesi, Antalya, Turkey}
\author{V. A. B. Zagatto}
\affiliation{Instituto de F\'isica, Universidade Federal Fluminense, Niter\'oi, Brazil}
\collaboration{for the NUMEN collaboration}
\noaffiliation
\date{\today}
\begin{abstract}
Single-nucleon transfer reactions are processes that selectively probe single-particle components of the populated many-body nuclear states. In this context, recent efforts are made to build a unified description of the rich nuclear spectroscopy accessible in heavy-ion collisions. An example of this multi-channel approach is the study of the competition between successive nucleon transfer and charge exchange reactions, the latter being of particular interest in the context of single and double beta decay studies. To this extent, the one-proton pickup reaction $^{48}$Ti($^{18}$O,$^{19}$F)$^{47}$Sc at 275 MeV was measured for the first time, under the NUMEN experimental campaign. Differential cross-section angular distribution measurements for the $^{19}$F ejectiles were performed at INFN-LNS in Catania by using the MAGNEX large acceptance magnetic spectrometer. The data were analyzed within the Distorted-Wave and Coupled-Channels Born Approximation frameworks. The initial and final-state interactions were described adopting the S\~ao Paulo potential, whereas the spectroscopic amplitudes for the projectile and target overlaps were derived from shell-model calculations. The theoretical cross-sections are found to be in very good agreement with the experimental data suggesting the validity of the optical potentials and the shell-model description of the involved nuclear states within the adopted model space.    
\end{abstract}
\maketitle
\vspace{0.2cm}
\vspace{0.3cm}
%
%
\section{Introduction}
In recent years, the interest of the physics community to study the neutrino-less double beta decay (0$\nu$$\beta$$\beta$ decay) \cite{iachello,vergados,shimizu,ejiri,decay_dol,liang} has been intensified although the aforementioned process has not been experimentally observed yet. Several experimental campaigns \cite{kamland2,nemo3,gerda3,candles,cuore,gerda2} have provided only lower limits on the 0$\nu$$\beta$$\beta$ decay half-life for selected $\beta$$\beta$ decay isotopes. The 0$\nu$$\beta$$\beta$ decay rate is governed by three factors namely, the phase-space factor G$_{0\nu}$ related to the motion of the electrons \cite{kotila}, the Nuclear Matrix Elements (NMEs) and a term containing the effective neutrino mass \cite{half_life}. Thus, the precise knowledge of the NMEs in conjunction with the experimental constraints for the decay half-life may provide the route towards the determination of the neutrino mass scale \cite{engel,ejiri}.\par
The NUMEN (NUclear Matrix Elements for Neutrinoless double beta decay) project \cite{numen_cappu} has been recently conceived at the Istituto Nazionale di Fisica Nucleare - Laboratori Nazionali del Sud (INFN-LNS) in Catania, Italy, aiming at accessing information about the NMEs of 0$\nu$$\beta$$\beta$ decay through the study of the heavy-ion induced double charge exchange (DCE) reactions on various $\beta$$\beta$ decay candidate targets. Among these, the $^{48}$Ti nucleus is under investigation since it is the daughter nucleus of $^{48}$Ca in the 0$\nu$$\beta$$\beta$ decay process \cite{48ca_beta_beta,abinitio_48ca}. The choice of DCE as surrogate reactions to study 0$\nu$$\beta$$\beta$ decay stems from the fact that both processes have several features in common. Among these, the two processes probe the same initial and final state nuclear wave functions \cite{numen_cappu,lenske,santopinto}, while Short-range Fermi, Gamow-Teller and rank-2 tensor components are present in both transition operators \cite{numen_cappu}. However, in order to extract meaningful information on the NMEs, contributions in the DCE channel from reaction mechanisms populating the same final nuclear states (based on direct meson-exchange or by sequential multi-nucleon transfer) should be quantitatively determined.\par
The degree of competition between DCE and multi-nucleon transfer is an important aspect for the description of the DCE reaction mechanism \cite{lenske,santopinto,bellone}. Early studies on the $^{40}$Ca($^{14}$C,$^{14}$O)$^{40}$Ar reaction at 51 MeV by Dasso and Vitturi showed an important contribution to the $^{40}$Ar$_{g.s.}$, at not very forward angles, from the sequential transfer of proton and neutron pairs \cite{dasso_dce_transfer}. Instead, the theoretical study of the DCE mechanism based on a recent measurement for the $^{40}$Ca($^{18}$O,$^{18}$Ne)$^{40}$Ar reaction at 275 MeV revealed that the direct-meson exchange mechanism may play a leading role, at least around zero degrees scattering angle \cite{cappu_pilot_exp,lay_dce_transfer}. However, the theoretical analysis in Ref.\cite{lay_dce_transfer} suggested the combination of single charge exchange (SCE) with one-proton and one-neutron transfer reactions as the second process in the leading order. Therefore, the contribution of the competitive processes to the DCE channel should not be taken for granted and measurements of all reaction channels under the same experimental conditions as the DCE reaction are necessary \cite{carbone_2p,cavallaro_2p,cavallaro_sce}.\par
Heavy-ion transfer reactions may provide valuable information on the nuclear structure and the reaction mechanism \cite{tr_mechanism,tr_mechanism3,morrison1,tr_mechanism2,tr_mechanism4,corradi,parmar,cardozo}. Due to their high selectivity in populating specific degrees of freedom in the residual nuclei, single-nucleon transfer are well established tools for probing single-particle configurations, while two-nucleon transfer offers an insight of pairing correlations \cite{scott,anyas,pairing,agodi}. One-proton transfer reactions have been extensively used for decades aiming at determining the spectroscopic factors. Experimentally, these quantities were traditionally determined by re-normalizing the calculated cross-sections, provided by a reaction model, to the measured ones (e.g. \cite{ursula,dwba_sf,dwba_sf2,dwba_sf3}). However, nowadays a substantial progress on reaction theory and in the computational and numerical sciences has been undertaken. The distortion of the incoming and outgoing scattering waves is under control by adopting double-folding optical potentials (OP) like the S\~ao Paulo Potential (SPP) \cite{sao_paulo1,sao_paulo2,sao_paulo3,sao_paulo4}. The available computational resources allow to perform exact finite-range calculations rather easily. In addition, with the unprecedented growth in computing power detailed nuclear structure calculations like large scale shell-model ones are increasingly adopted for the determination of spectroscopic factors. Into this context, it has been demonstrated that it is possible to obtain reliable spectroscopic information from heavy-ion transfer reactions without the need for any arbitrary scaling factor \cite{caval,cappu_nature,ermamatov,carbone2,ermamatov2,santagati,linares_1n,zagatto2,linares2}.\par
One of the most useful reaction models for the analysis of experimental data on transfer reactions is the Distorted-Wave Born Approximation (DWBA) formalism \cite{satchler,timofeyuk}. The main ingredients of the distorted-wave theory are the OP, which describe the elastic scattering at the entrance and exit channels, and the overlap functions which contain information on the nuclear structure and angular momentum of the involved nuclei. At energies above the Coulomb barrier, the DWBA calculation is rather sensitive on the choice of the OP parameters \cite{dwba_sf,op_effect}. On the other hand, the overlap functions are usually determined as single-particle solutions of a Woods-Saxon potential weighted by the corresponding spectroscopic amplitudes provided by many-body shell model calculations. In a recent publication of our group, the importance of these model dependencies for the description of two-nucleon transfer for $^{20}$Ne+$^{116}$Cd system was addressed \cite{carbone_2p}.\par
The success of the DWBA to describe one-nucleon transfer data is well-established in the literature \cite{montanari,gasques} but in some cases, when coupling effects among the various reaction channels are strong, it fails to reproduce the experimental data. In particular, projectile and/or target excitation could play a deep role in the reaction mechanism and thus should be included in the theoretical description by means of the Coupled-Channels Born Approximation (CCBA) method \cite{ccba,ccba_effect,keeley}.\par 
The $^{18}$O has been frequently used as projectile in several studies, but primarily involving two-neutron transfer reactions \cite{bernas,sahu,bondi,caval,cappu_nature,ermamatov,carbone2,ermamatov2,santagati}. This stems from the fact that the neutron pair out of the $^{16}$O core can be transferred to another nucleus during a nuclear collision thus giving the opportunity to study neutron pairing correlations in the final nuclear states. On the other hand, proton transfer reactions has not been thoroughly investigated. The ($^{18}$O,$^{19}$F) reaction on $^{40}$Ca and $^{54}$Fe has been studied in Ref.\cite{18o_siemssen} yielding poor results in the description of the angular distributions. In a more recent publication \cite{zagatto2} the same reaction was considered in a systematic study together with one-neutron, two-neutron and $\alpha$-particle transfer for the system $^{18}$O+$^{65}$Cu, in order to investigate the effect of transfer reactions on the quasi-elastic barrier distribution.\par
In the present work, the one-proton pickup reaction for the system $^{18}$O+$^{48}$Ti at 275 MeV incident energy is investigated for the first time, under the NUMEN experimental campaign. The measured differential cross-section angular distribution data are analyzed within the DWBA framework in order to validate the adopted optical potentials for the description of the initial and final-state interactions as well as to access information on the single-particle components of nuclear wave functions of the involved nuclei. Further on, the coupling influence of projectile and target excitations are investigated adopting the CCBA formalism. The data analysis is also accompanied by the study of the same reaction on $^{16}$O and $^{27}$Al targets, measured under the same experimental conditions, in order to estimate the background arising from the different target components (TiO$_2$+$^{27}$Al) as well as to strengthen the conclusions of our analysis. As a result of this work we highlight the importance of studying transfer reactions, since they provide the ground for testing the validity of the initial and final-state interactions and the spectroscopic information provided by the nuclear structure models, which should be under control for the proper description of the SCE and DCE mechanisms \cite{lenske}.\par
This paper is organized as follows: the experimental setup and data reduction procedure are reported in Sections II and III, respectively; a brief description of the theoretical formalism used for the calculation of the one-proton transfer cross-sections is given in Section IV; our results are discussed in Section V and the conclusions are presented in Section VI.
\section{Experimental setup}
The experiment was carried out at INFN-LNS in Catania. The $^{18}$O$^{8+}$ ion beam, delivered by the K800 Superconducting Cyclotron at the energy of 275 MeV, was directed onto a 510 $\mu$g/cm$^2$ thick TiO$_2$ target which was evaporated on a 216 $\mu$g/cm$^2$ thick aluminum foil. Measurements using a WO$_3$ target (284 $\mu$g/cm$^2$) evaporated on an aluminum foil and a 226 $\mu$g/cm$^2$ thick $^{27}$Al one were performed, to estimate the background arising from the different target components. The use of a collimation system composed of a 2 mm diameter collimator followed by a 4 mm diameter anti-scattering, mounted 125 mm and 20 mm upstream the target respectively, limited the beam spot size to $\approx$3 mm at the target position and the beam angular divergence to 3 mrad. The beam charge was collected by a Faraday cup located 150 mm downstream the target position. The error in the measurement of the beam charge was reduced by placing at the entrance of the Faraday cup an electron suppression ring biased at -200V, minimizing the escape of the secondary produced electrons.\par
The various reaction products were momentum analyzed by the MAGNEX large acceptance magnetic spectrometer \cite{magnex}, whose optical axis was set at $\theta$$_{opt}$= \ang{9} with respect to the beam direction. MAGNEX was operated in full horizontal acceptance covering an angular range between \ang{3} and \ang{15} in the laboratory reference frame but with a reduced vertical one ($\pm$\ang{2}), in order to avoid the high ion counting rate at the Focal Plane Detector (FPD) \cite{fpd,fpd2,fpd3}. The latter was used in order to identify the ions of interest among the various reaction products, by means of the conventional $\Delta$E-E method for the Z separation and a technique for the determination of the mass number based on the correlation between the ions kinetic energy and the measured horizontal position at the focal plane. An example of identification spectra is presented in Fig.1, while a detailed description of the identification technique is given in Ref.\cite{pid}.
%
%
%
%
\begin{figure}[!h]
\begin{center}
\includegraphics[width=0.50\textwidth]{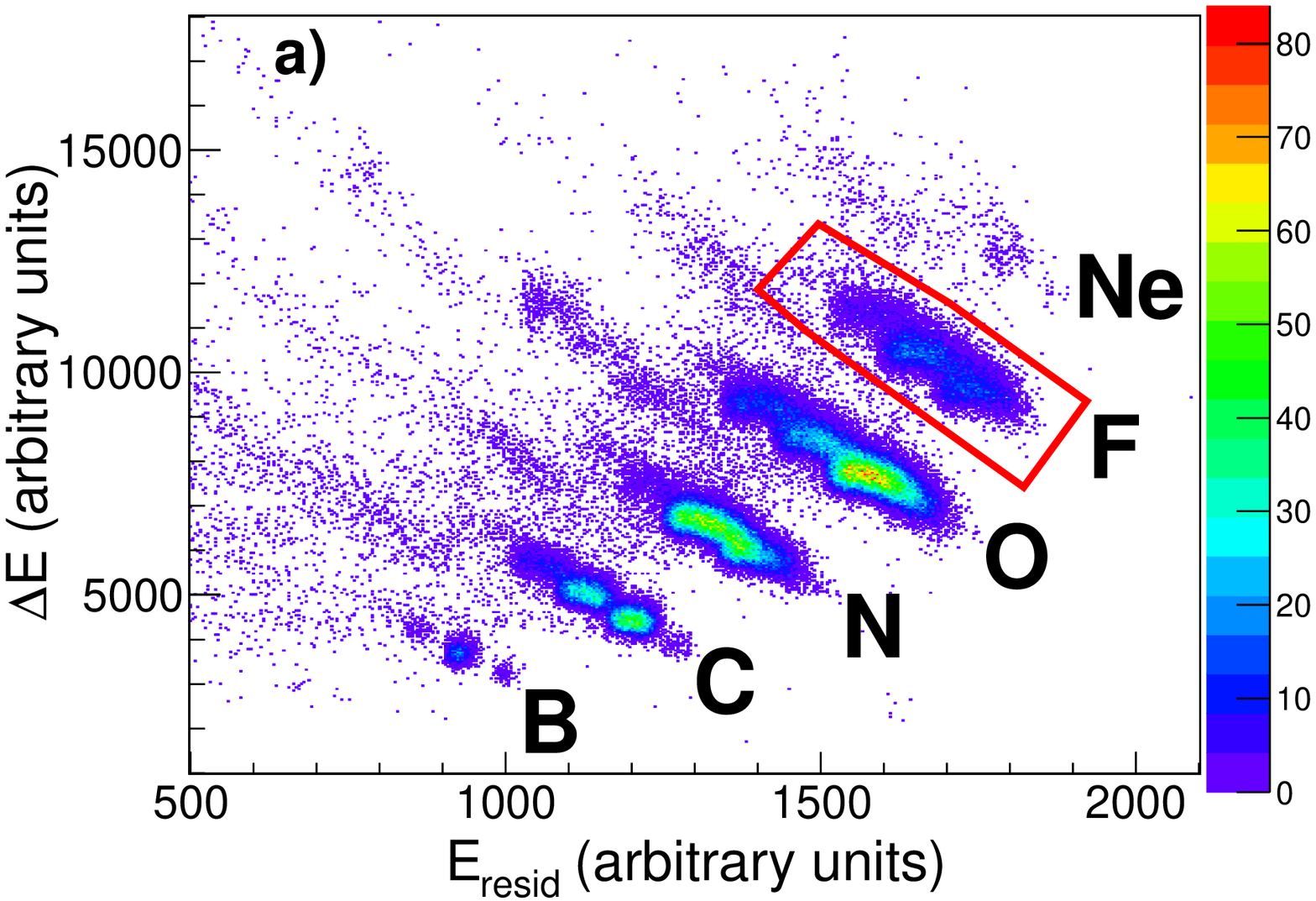}
\vspace{0.1cm}
\includegraphics[width=0.50\textwidth]{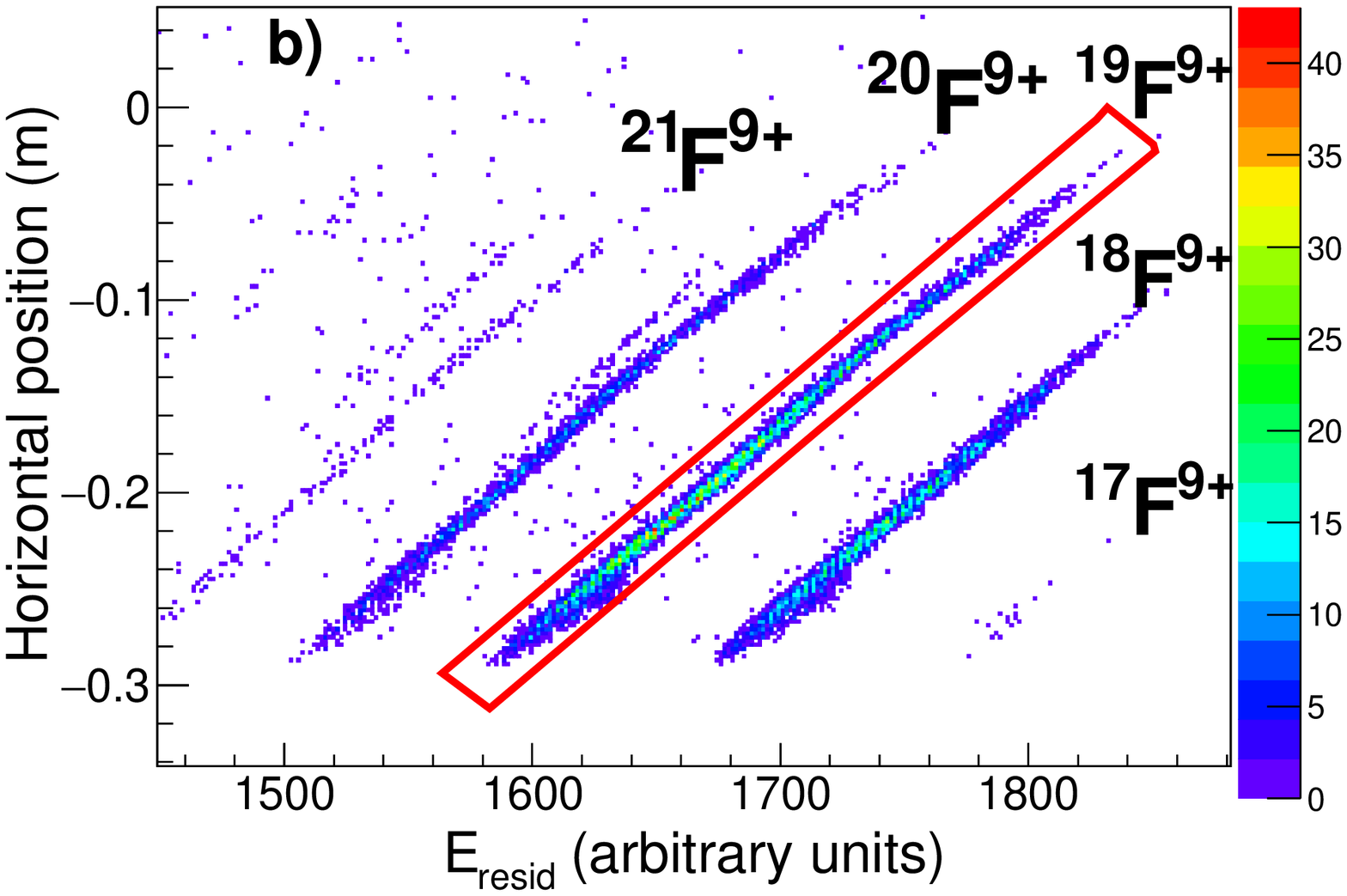}
\vspace{0.1cm}
\caption{Typical identification spectra for $^{48}$Ti($^{18}$O,$^{19}$F)$^{47}$Sc reaction at the energy of 275 MeV. a) $\Delta$E-E$_{resid}$ correlation plot for the ions energy loss inside the gas tracker, $\Delta$E, as a function the of residual energy, E$_{resid}$, measured by one silicon detector of the FPD. A graphical selection in the contour of the fluorine ion events is illustrated by the solid red line. b) The horizontal position at the MAGNEX focal plane as a function of the residual energy for the selected fluorine ions of panel (a). The different loci correspond to ions with different ratio $\sqrt m$/q. A graphical selection on the $^{19}$F$^{9+}$ events is shown by the solid red line.}
\end{center}
\end{figure}
\section{Data reduction}
Once the $^{19}$F ejectiles were identified in the spectra, a high-order software ray reconstruction technique was applied to the data by solving the equation of motion for the ejectile particles, based on a fully differential algebraic method. Hence, it was possible to reconstruct the ion momentum vector at the target position \cite{momentum_vector}. The reconstruction procedure was performed separately for each of the reactions under study namely, the $^{48}$Ti($^{18}$O,$^{19}$F)$^{47}$Sc, $^{16}$O($^{18}$O,$^{19}$F)$^{15}$N and the $^{27}$Al($^{18}$O,$^{19}$F)$^{26}$Mg reaction corresponding to the measurement with the TiO$_2$, WO$_3$ and $^{27}$Al targets, respectively.
\subsection{The $^{27}$Al($^{18}$O,$^{19}$F)$^{26}$Mg and $^{16}$O($^{18}$O,$^{19}$F)$^{15}$N reactions}
The first part of the data reduction is referred to the analysis of the one-proton transfer events obtained with the aluminum target, since the aluminum backing was present in both measurements performed with the TiO$_2$ and WO$_3$ targets. The absolute cross-sections were determined after correcting the experimental yields for the overall efficiency of MAGNEX spectrometer \cite{effic}. In Fig.2a the absolute differential cross-section as a function of the excitation energy for the $^{27}$Al($^{18}$O,$^{19}$F)$^{26}$Mg reaction corresponding to the angular range between \ang{3.5} and \ang{14} in the laboratory reference frame is shown. The excitation energy E$_{x}$ was obtained from the missing mass method \cite{magnex} as:
%
%
\begin{equation}
E_{x} = Q_{0}-Q,
\end{equation} 
where Q$_{0}$ is the ground state (g.s.) to g.s. Q-value for the $^{27}$Al($^{18}$O,$^{19}$F)$^{26}$Mg reaction and Q is a term containing the masses of the nuclei at the entrance and exit channels, and the reconstructed kinetic energy and angle of the $^{19}$F ions. The obtained energy resolution in the current measurement was approximately 500 keV in full width at half maximum (FWHM). Looking at Fig.2a, various states are populated up to $\alpha$-particle emission threshold of $^{26}$Mg (S$_{\alpha}$$\approx$ 10.6 MeV), while a rather continuous shape is observed beyond such threshold. In order to describe the excitation energy spectrum in terms of the well-known states of the $^{19}$F and $^{26}$Mg nuclei a fit procedure was performed, where the shape of each state was described by a Gaussian function. The results of the fit procedure are presented in Fig.2b. It should be pointed out that the output of the fit is the overall contribution of the Gaussian peaks, rather than the individual contribution of each transition. Therefore, the experimental yields of the unresolved states corresponding to three different excitation energy regions namely, -1.0 $<$ E$_x$ $<$ 1.0 MeV, 1.0 $<$ E$_x$ $<$ 3.0 MeV and 3.5 $<$ E$_x$ $<$ 5.5 MeV were deduced from the fits considering an angular step between \ang{0.5} and \ang{4} depending on the statistics. The differential cross-section angular distributions were extracted and are presented in Fig.3. The error bars include the statistical uncertainty and to a lesser extent a contribution due to the uncertainty in the determination of the solid angle. A systematic error of about 10 \% due to the uncertainty in the target thickness and the integrated value of the beam charge, common to all the data points, is not included in the error bars.\par
%
%
%
\begin{figure}[!h]
\begin{center}
\includegraphics[width=0.50\textwidth]{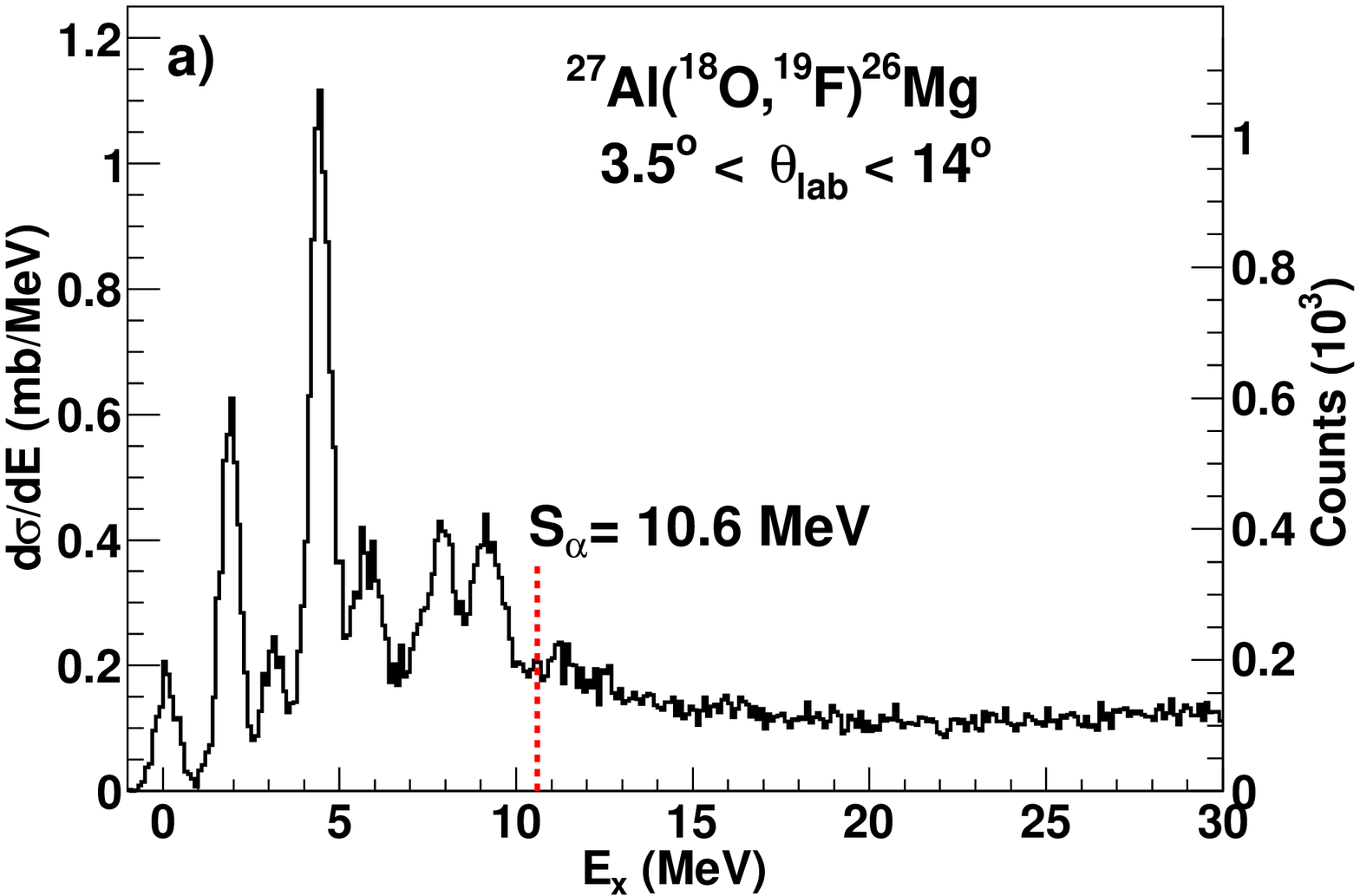}
\includegraphics[width=0.50\textwidth]{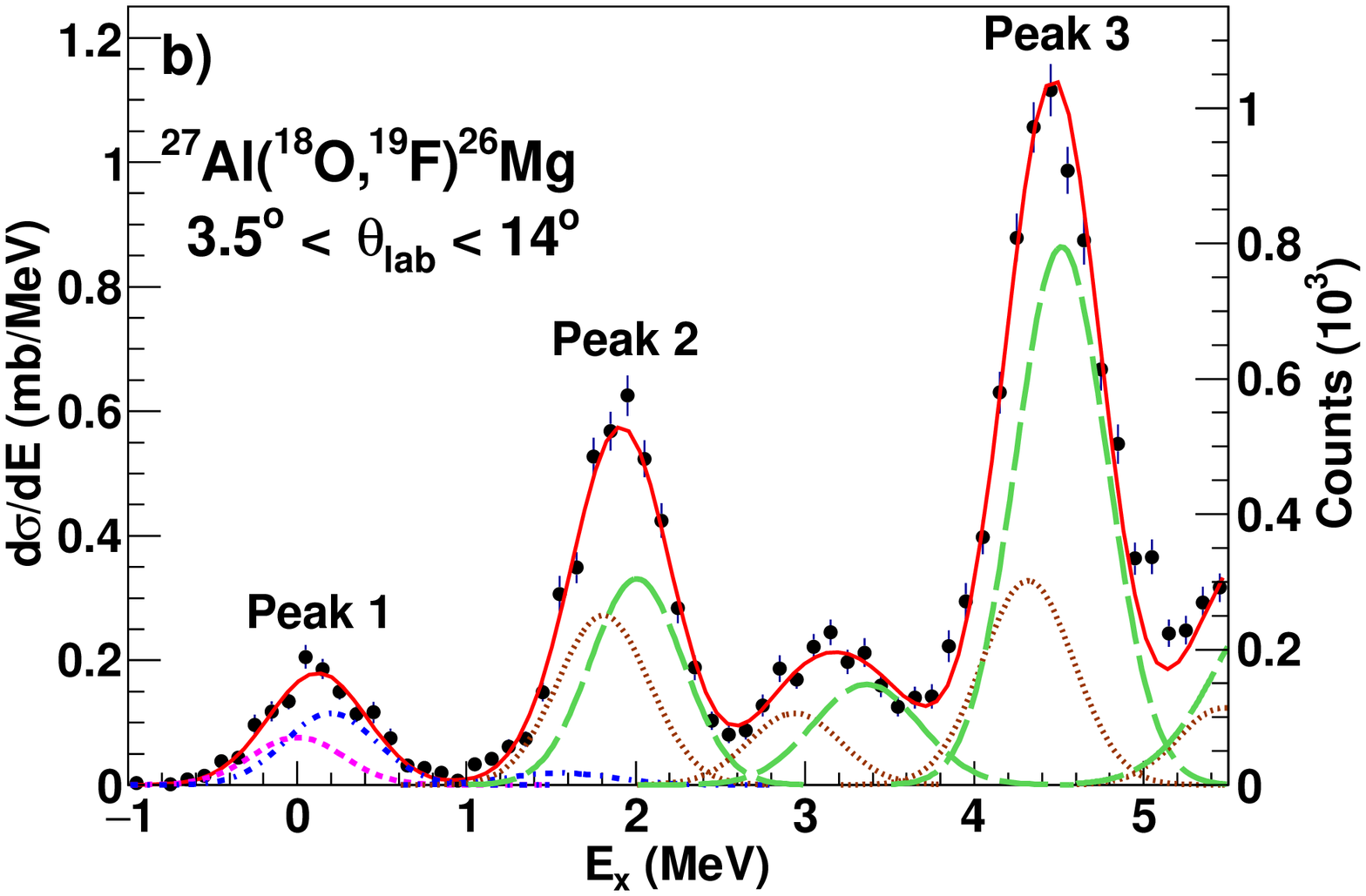}
\caption{a) Reconstructed excitation energy spectrum for the $^{27}$Al($^{18}$O,$^{19}$F)$^{26}$Mg reaction at 275 MeV. The alpha emission threshold of $^{26}$Mg nucleus is indicated by the vertical red dashed line. b) Same as in panel (a), but with a zoom of the first 5 MeV of the excitation energy spectrum. The experimental data are compared to the result of a fit procedure where for the description of each known state a Gaussian function was used. The dashed magenta curve corresponds to the ground state to ground state transition, the dotted-dashed blue lines are referred to the excited states of the $^{19}$F nucleus, the dotted brown lines correspond to the excited states of $^{26}$Mg keeping the $^{19}$F in its ground state and the long-dashed green lines correspond to transitions where both ejectile and residual nuclei are excited. The red line corresponds to sum of the Gaussian curves.}
\end{center}
\end{figure}
%
%
%
\begin{figure}
\begin{center}
\includegraphics[width=0.52\textwidth]{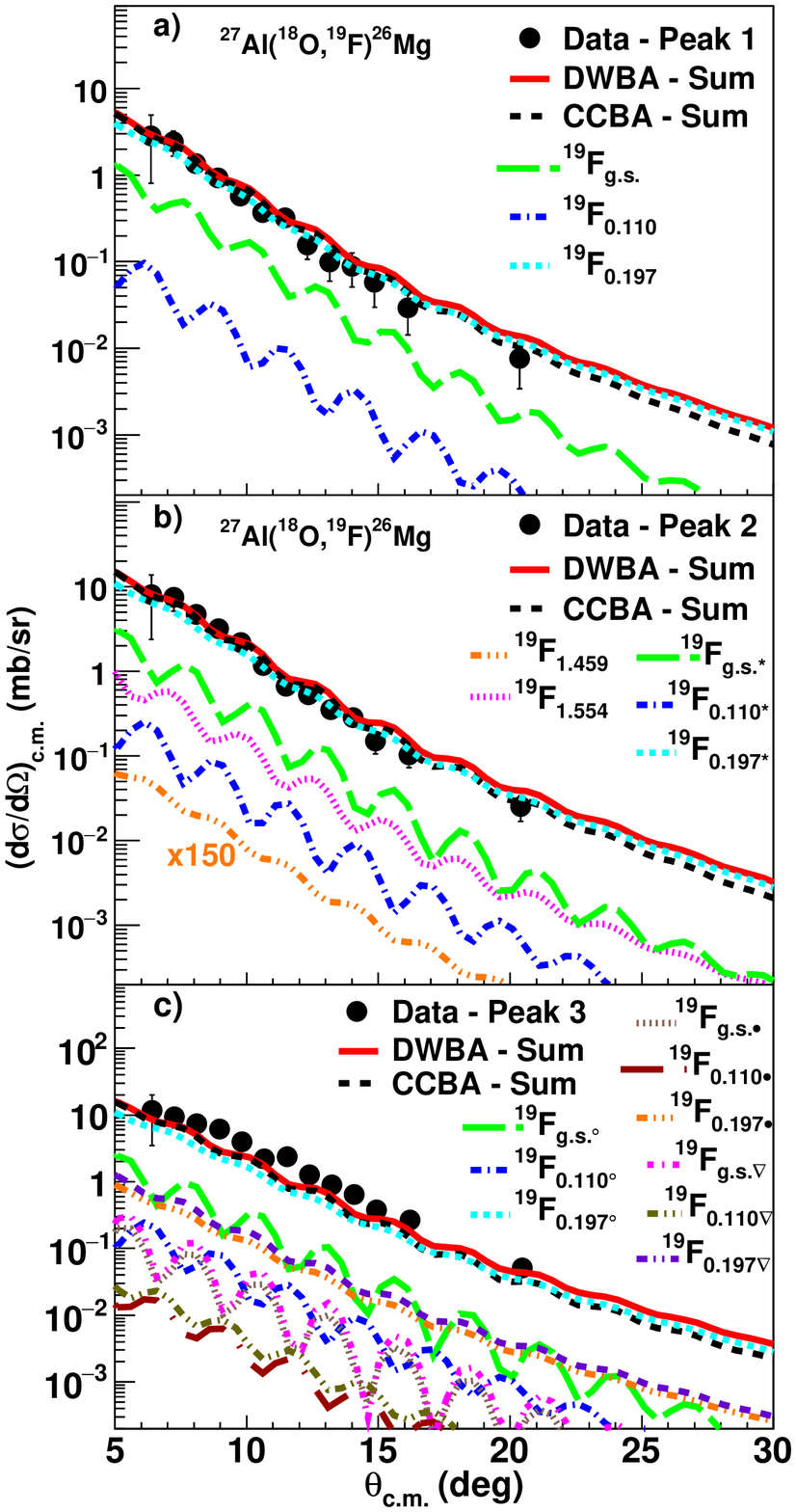}
\caption{Comparison between experimental and theoretical angular distribution data for the $^{27}$Al($^{18}$O,$^{19}$F)$^{26}$Mg reaction measured at 275 MeV. The experimental data, denoted with the black circles, were obtained by integrating the yields of the unresolved states referred as a) first peak, b) second peak and c) third peak in Fig.2. Theoretical cross-sections for the transitions to the involved states of the ejectile and the residual nuclei were calculated within the DWBA framework and are presented with the colored curves. In the legend, each curve is labeled by the corresponding excitation energy of $^{19}$F for transitions to the $^{26}$Mg$_{g.s.}$ and a symbol in case where $^{26}$Mg is excited. The curves marked with an asterisk correspond to transitions to the \(2\)$_{1}$$^{+}$ (1.809 MeV) state of $^{26}$Mg, while those marked with an open circle, a filled circle and a triangle correspond to transitions to the \(4\)$_{1}$$^{+}$ (4.319 MeV), \(2\)$_{3}$$^{+}$ (4.333 MeV) and \(3\)$_{3}$$^{+}$ (4.350 MeV) states of $^{26}$Mg, respectively. The sum of all transitions is illustrated by the red solid line. The sum of a CCBA calculation considering the same final states as the DWBA one is indicated with the dashed black curve.}
\end{center}
\end{figure}
As stated above, to study the $^{16}$O($^{18}$O,$^{19}$F)$^{15}$N reaction, a compound target of WO$_3$ evaporated onto a thin aluminum foil was used. Therefore, the background due to the presence of the two contaminants should be identified in order to isolate the $^{19}$F and $^{15}$N spectra. The reconstructed excitation energy spectrum measured in the angular range between \ang{4} and \ang{14} in the laboratory reference frame is shown in Fig.4. The excitation energy was determined using Eq.1 with Q$_{0}$ being the g.s. to g.s Q-value of the $^{16}$O($^{18}$O,$^{19}$F)$^{15}$N reaction. The pronounced peaks at E$_{x}$ $\approx$ -5 MeV and -3 MeV were identified through the reaction kinematics with the one-proton transfer reaction on $^{27}$Al. Thus, by using the data obtained with the pure aluminum target it was possible to subtract the aluminum contribution, appropriately normalized, from the spectrum. A small remnant of events located at E$_{x}$$<$ -8 MeV was attributed to the tungsten component of the target. In the absence of a measurement with a pure tungsten target, a uniform distribution for the tungsten contamination was assumed throughout the whole energy range (see Fig.4). Subsequently, the tungsten contribution was subtracted and the excitation energy spectrum for the $^{18}$O + $^{16}$O reaction was deduced. Following the same procedure as the one adopted for the $^{18}$O + $^{27}$Al reaction, energy and angle differential cross-sections were deduced and are presented in Figs. 5 and 6, respectively. The error in both spectra includes the contribution from the statistical and the background subtraction uncertainties. Having an inspection at the energy spectrum of Fig.5, it is seen that the first excited state of the $^{15}$N nucleus is found at E$_{x}$= 5.27 MeV. So, the observed structure at the first $\approx$ 5 MeV of the spectrum in Fig.5 is the fingerprint of the population of $^{19}$F low-lying states. This offers the unique possibility for validating the spectroscopic information for the $^{19}$F states provided by our shell-model calculations and thus to optimize the overall data interpretation. Further details are given in the following section.   
%
%
%
%
\begin{figure}
\begin{center}
\includegraphics[width=0.50\textwidth]{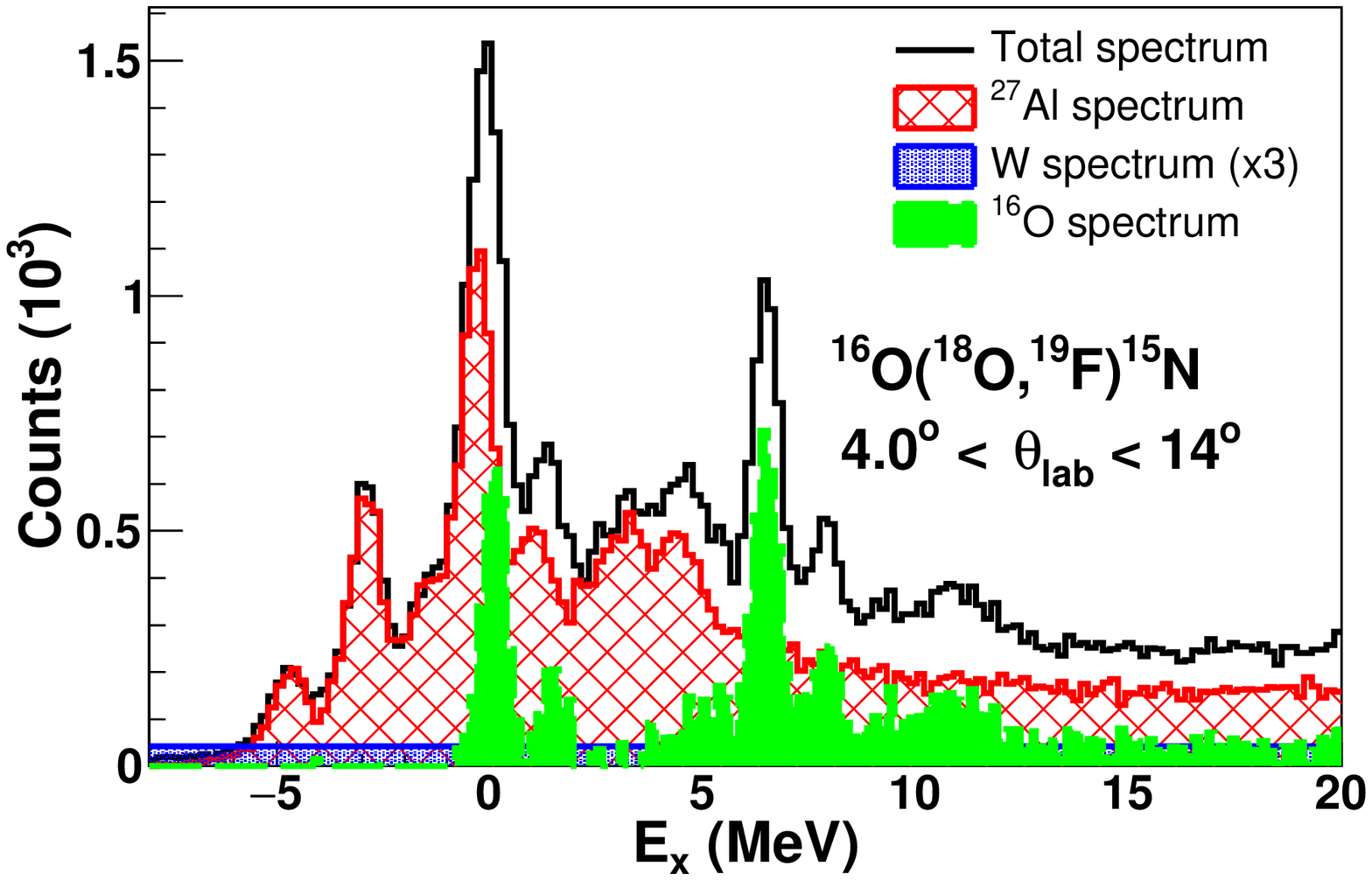}
\caption{Decomposition of the excitation energy spectrum obtained with the WO$_{3}$+$^{27}$Al target. The total spectrum is presented with the solid black line. The red-hatched area corresponds to the normalized background originating from the $^{27}$Al backing material, while the dotted blue area corresponds to the background arising from tungsten component of the target. For reasons of clarity, the latter is multiplied by a factor of 3. The solid green area is the obtained excitation energy spectrum for the $^{16}$O($^{18}$O,$^{19}$F)$^{15}$N reaction, after subtracting from the total spectrum the background contributions.}
\end{center}
\end{figure}
%
%
%
%
\begin{figure}[!h]
\begin{center}
\includegraphics[width=0.50\textwidth]{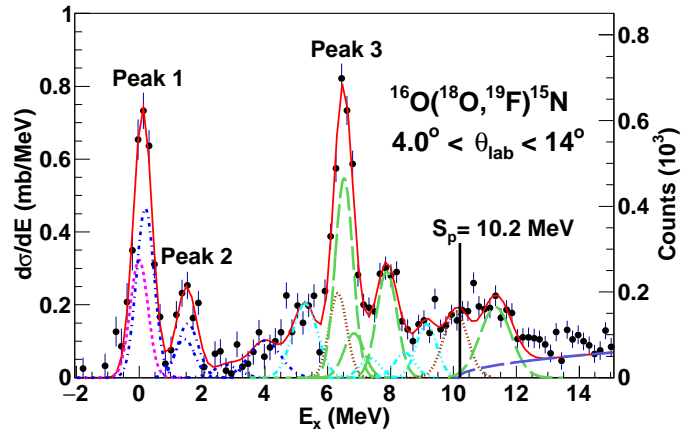}
\caption{Excitation energy spectrum for the $^{16}$O($^{18}$O,$^{19}$F)$^{15}$N reaction at the energy of 275 MeV compared to the results of the fit procedure, where for the description of each state, a Gaussian function was used. The adopted colors and line styles are the same as those presented in Fig.2. In addition to Fig.2, in case of several nearby states a single Gaussian function, depicted with the double-dotted-dashed curve, was considered in the fit procedure with the centroid fixed to the mean value of the excited states. Moreover, the continuum background above the proton emission threshold of $^{15}$N is indicated with the purple line.}
\end{center}
\end{figure}
%
%
%
%
\begin{figure}
\begin{center}
\includegraphics[width=0.52\textwidth]{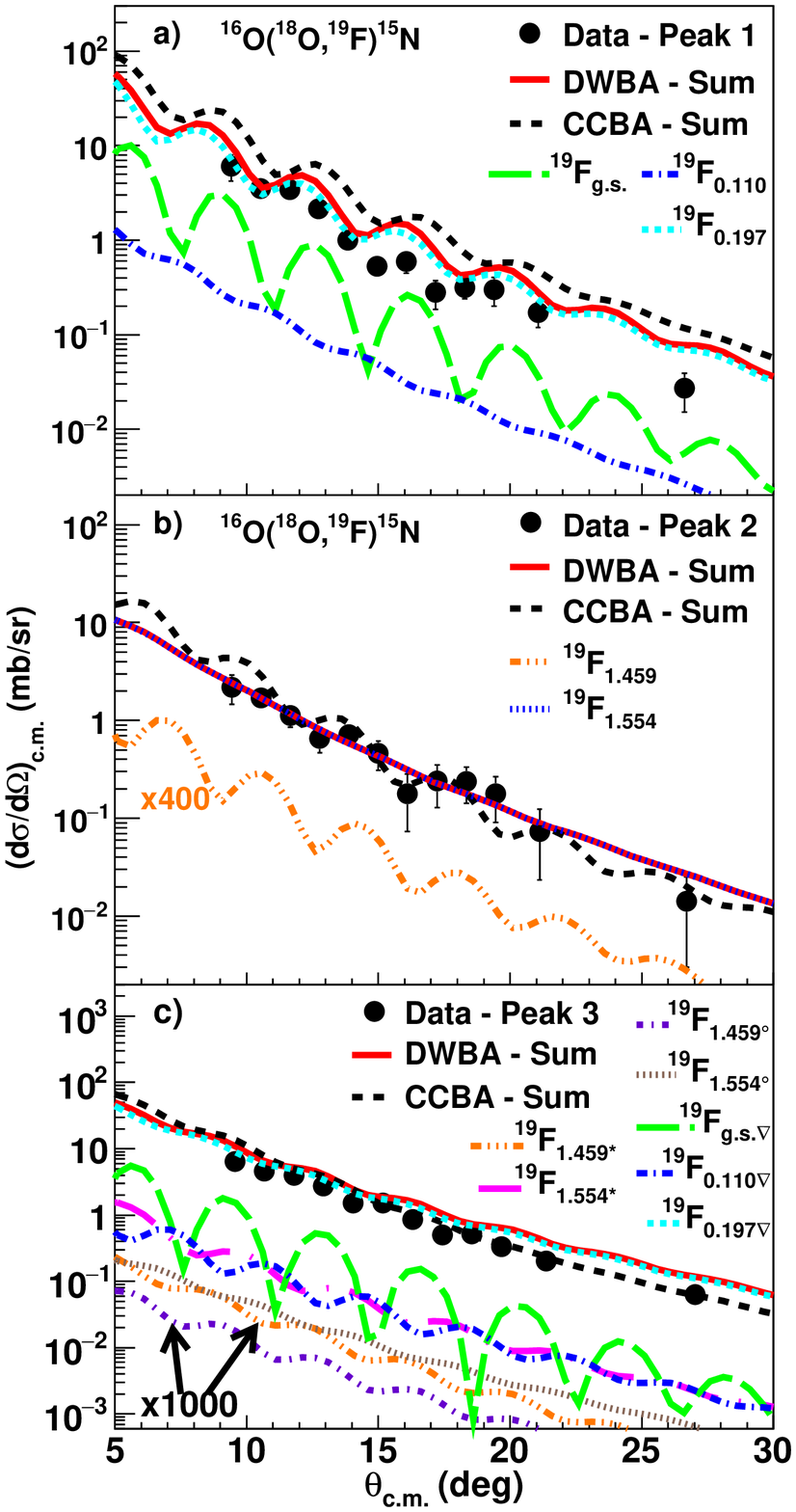}
\caption{Comparison between experimental and theoretical angular distribution data for the $^{16}$O($^{18}$O,$^{19}$F)$^{15}$N reaction measured at 275 MeV. The experimental data, denoted with the black circles, were obtained by integrating the yields of the unresolved states referred as a) first peak, b) second peak and c) third peak in Fig.5. Theoretical angular distributions for the transitions to the involved states of the ejectile and the residual nuclei were calculated within the DWBA framework and are presented with the colored curves. In the legend, each curve is labeled by the corresponding excitation energy of $^{19}$F for transitions to the $^{15}$N$_{g.s.}$ and a symbol in case where $^{15}$N is excited. The curves marked with an asterisk, an open circle and a triangle correspond to transitions to the \(\frac{5}{2}\)$_{1}$$^{+}$ (5.27 MeV), \(\frac{1}{2}\)$_{1}$$^{+}$ (5.299 MeV) and \(\frac{3}{2}\)$_{1}$$^{-}$ (6.324 MeV) states of $^{15}$N, respectively. The sum of all transitions is illustrated by the red solid line. The sum of a CCBA calculation considering the same final states as the DWBA one is indicated with the dashed black curve.}
\end{center}
\end{figure}
\subsection{ The $^{48}$Ti($^{18}$O,$^{19}$F)$^{47}$Sc reaction}
The reconstructed excitation energy spectrum obtained with the TiO$_2$+$^{27}$Al target is shown in Fig.7. The excitation energy was obtained through Eq.1 with Q$_{0}$ being the g.s. to g.s Q-value of the $^{48}$Ti($^{18}$O,$^{19}$F)$^{47}$Sc reaction. In the same figure, the contributions from the $^{27}$Al backing and $^{16}$O component of the target, appropriately normalized, are also shown. The background spectra were subtracted from the total one and the excitation energy spectrum for the $^{48}$Ti($^{18}$O,$^{19}$F)$^{47}$Sc reaction was deduced. Unlike spectra in Figs. 2 and 5, this one has a rather continuum shape, reflecting the high level density of the populated $^{47}$Sc nucleus and the limited energy resolution. Therefore, angular distributions were determined by integrating the events in a wide energy range corresponding to 0 $<$ E$_x$ $<$ 3.5 MeV including thus the contribution of various excited states of $^{19}$F and $^{47}$Sc nuclei. The obtained angular distribution data are presented in Fig.8. 
%
%
%
\begin{figure}
\begin{center}
\includegraphics[width=0.50\textwidth]{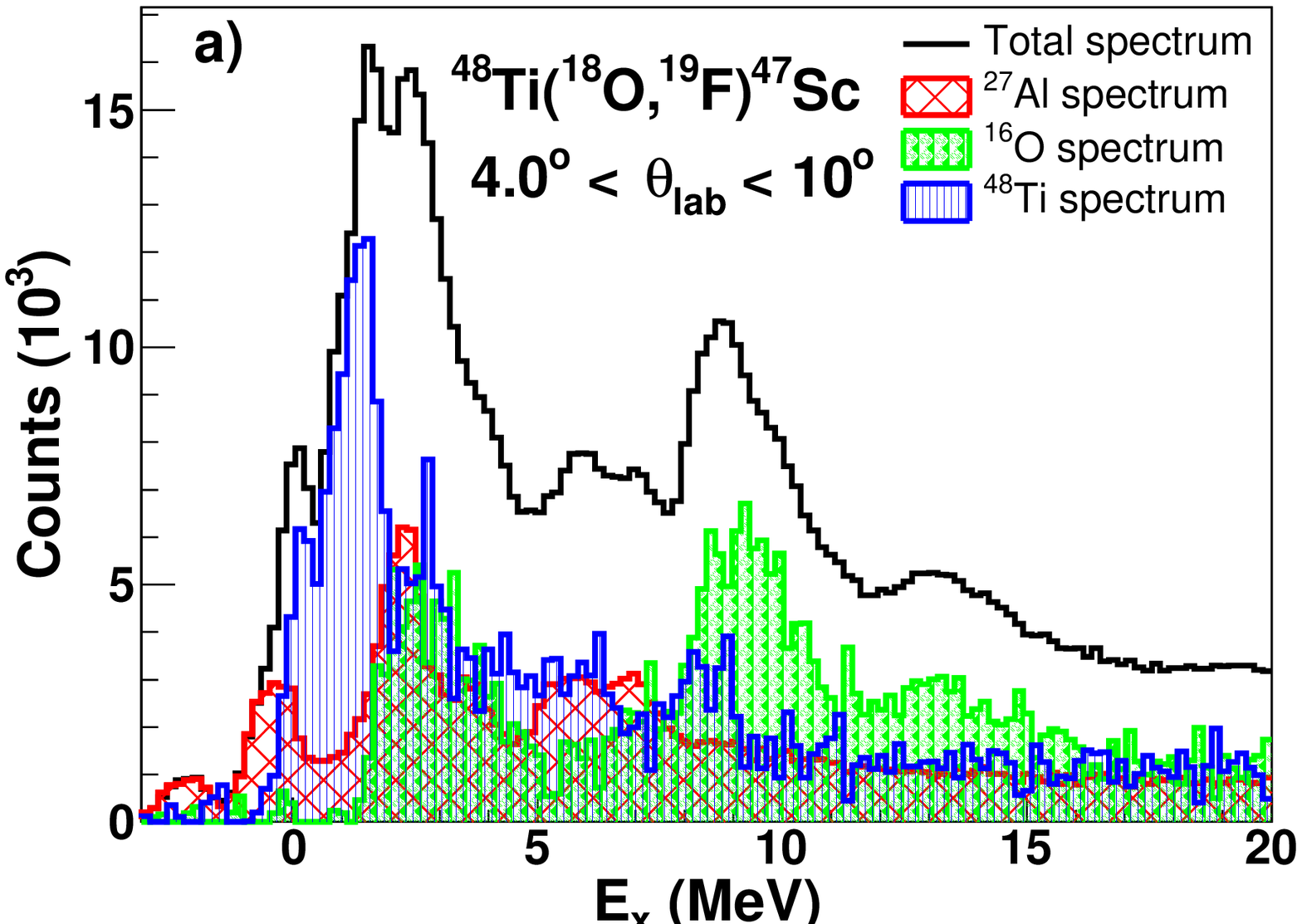}
\includegraphics[width=0.50\textwidth]{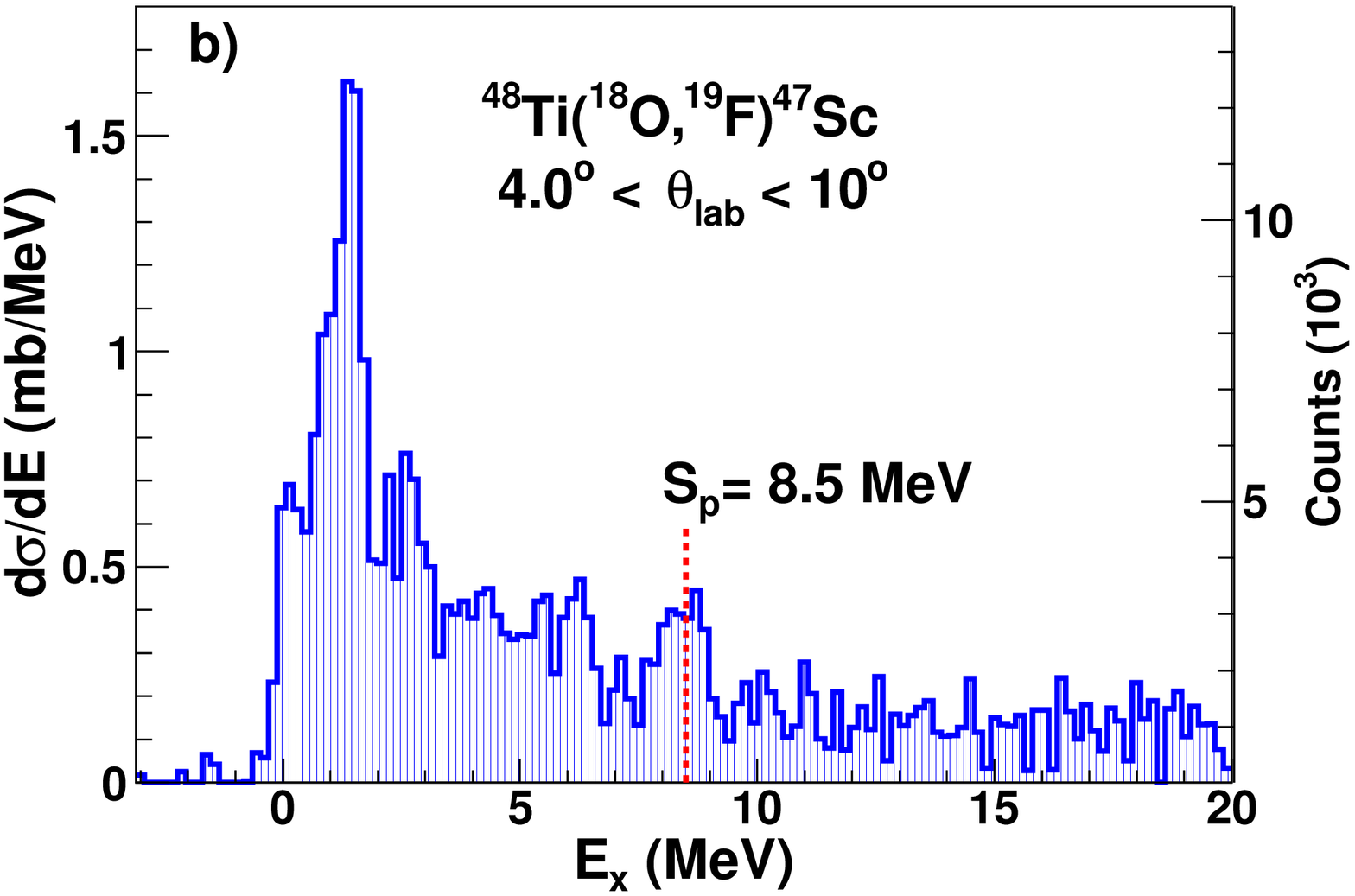}
\caption{a) Decomposition of the excitation energy spectrum obtained with the TiO$_{2}$+$^{27}$Al target. The total spectrum is presented with the solid black line. The red-hatched area corresponds to the normalized background originating from the $^{27}$Al backing material, while the dotted green area corresponds to the background arising from oxygen. The blue-vertically-hatched  area is the obtained excitation energy spectrum for the $^{48}$Ti($^{18}$O,$^{19}$F)$^{47}$Sc reaction, after subtracting from the total spectrum the background contributions. b) Reconstructed excitation energy spectrum for the $^{48}$Ti($^{18}$O,$^{19}$F)$^{47}$Sc reaction at 275 MeV. The proton emission threshold of $^{47}$Sc nucleus is indicated with the vertical red dashed line.}
\end{center}
\end{figure}
%
%
%
\begin{figure}
\begin{center}
\includegraphics[width=0.50\textwidth]{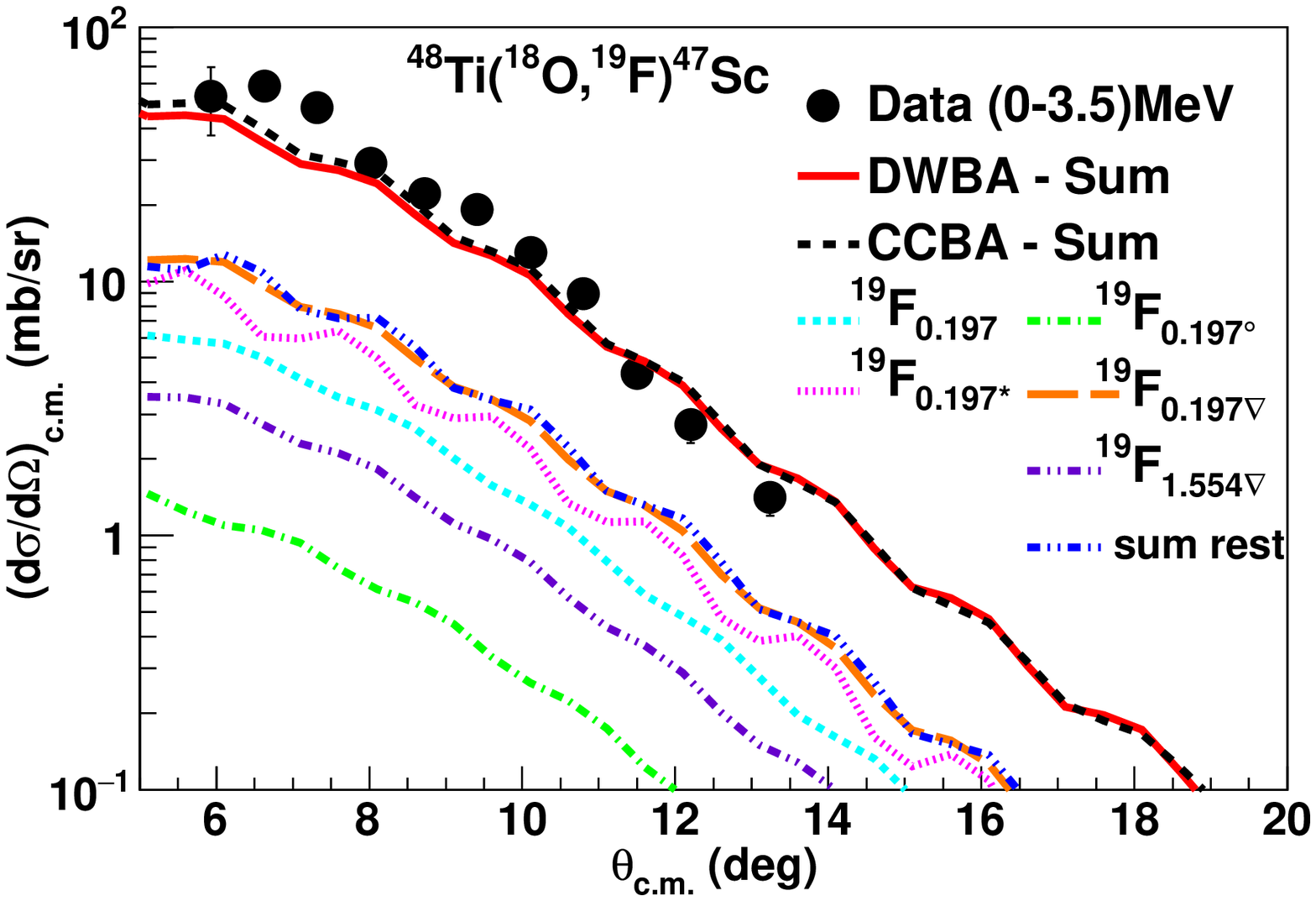}
\caption{Comparison between experimental and theoretical cross-sections for the $^{48}$Ti($^{18}$O,$^{19}$F)$^{47}$Sc reaction measured at 275 MeV. The experimental data, denoted with the black circles, were obtained by integrating the spectrum of Fig.7 between 0 $<$ E$_x$ $<$ 3.5 MeV. Theoretical angular distributions for the transitions to the involved states of the ejectile and the residual nuclei were calculated within the DWBA framework and are presented with the colored curves. Due to the large number of states introduced in the calculation, for reasons of clarity only the five stronger transitions are shown in the graph. In the legend, each curve is labeled by the corresponding excitation energy of $^{19}$F for transitions to the $^{47}$Sc$_{g.s.}$ and a symbol in case where $^{47}$Sc is excited. The curves marked with an asterisk, an open circle and a triangle correspond to excitation in the \(\frac{3}{2}\)$_{1}$$^{+}$ (0.767 MeV), \(\frac{3}{2}\)$_{1}$$^{-}$ (0.808 MeV) and \(\frac{1}{2}\)$_{1}$$^{+}$ (1.391 MeV) states of $^{47}$Sc respectively. The contribution of all other transitions is indicated with the blue dotted-dashed line under the notation "sum rest". The sum of all transitions is illustrated by the red solid line. The sum of a CCBA calculation considering the same final states as the DWBA one is indicated with the dashed black curve.}
\end{center}
\end{figure}
\section{Theoretical analysis}
The experimental angular distribution data were analyzed within the DWBA and CCBA frameworks using FRESCO code \cite{fresco}. Considering a reaction of the form: a + A $\rightarrow$ b + B, the transition amplitude in the DWBA model is given by the following expression:
%
%
\begin{equation}
T = \int  d\vec{r_{\alpha}} d\vec{r_{\beta}}  \chi_{\beta}^{(-)*} \braket{\Psi_{b} \Psi_{B}|V|\Psi_{a} \Psi_{A}}  \chi_{\alpha}^{(+)},
\end{equation} 
where $\chi$$_{\alpha}$ and $\chi$$_{\beta}$ are the distorted waves describing the relative motion of the involved nuclei at the entrance ($\alpha$) and exit ($\beta$) channels, respectively and $\braket{\Psi_{b} \Psi_{B}|V|\Psi_{a} \Psi_{A}}$ is the matrix element describing the interaction between the internal states of the colliding pairs. The transfer operator V was calculated in the post representation including full complex remnant terms. In the present analysis, for all reactions under study the distorted waves at the entrance and exit channels were generated adopting the double-folding S\~ao Paulo Potential for the description of both the real and imaginary parts of the optical potential, but with two different normalization factors N$_R$ and N$_I$, respectively. For the DWBA calculations, the adopted values for the normalization factors were N$_R$=1.0 and N$_I$= 0.78. Based on a systematic study, it was found that with this prescription the SPP is able to describe adequately-well the elastic scattering data for heavy-ion reactions involving light \cite{spp_light1,spp_light2}, medium \cite{spp_med1,spp_med2} and heavy \cite{spp_heav1,carbone_universe} mass targets over a wide energy range. In the CCBA approach where couplings to the relevant projectile and target excitations are explicitly taken into account, the normalization factor of the imaginary part of the OP at the entrance channel was reduced to N$_I$= 0.60. The same prescription has been also adopted in previous studies for the description of one and two-neutron transfer reactions \cite{sao_paulo4,linares_1n,caval,ermamatov,carbone2,ermamatov2}. It should be noted that for the $^{27}$Al($^{18}$O,$^{19}$F)$^{26}$Mg reaction we considered only couplings to the \(2\)$_{1}$$^{+}$ state of $^{18}$O and thus an intermediate value for the normalization factor N$_I$= 0.70 was chosen.\par
The single-particle wave functions were calculated assuming that the transferred proton is bound to the core by an effective potential of a Woods-Saxon form factor with a radius R= r$_{0}$A$_{i}$$^{1/3}$, where A$_{i}$ is the mass number of the core nucleus. For the $^{18}$O core the reduced radius (r$_0$) and diffuseness was set to 1.26 and 0.70 fm, respectively, whereas the reduced radii and the diffusenesses were set to 1.20 and 0.60 fm, respectively, for the heavier nuclei \cite{carbone_2p,cavallaro_2p}. For the potential that binds the valence proton in the $^{15}$N core, since $^{15}$N has an excess of one neutron with respect to protons, the adopted geometry of the binding potential was the same as for the $^{18}$O neutron rich core. Tests adopting different values for the reduced radius (1.20-1.26) fm and diffuseness (0.60-0.70) fm of this potential were performed showing that the theoretical cross-sections for one-proton transfer do not vary significantly. In all cases, the depth of the potential was adjusted such as to reproduce the separation energy of the transferred particle.\par
In the CCBA calculation, inelastic excitations to the low-lying states of the projectile and target nuclei were taken into account adopting the rotational model. Coulomb deformations were introduced in terms of the reduced transition probabilities B(E$_{\lambda}$), where $\lambda$ is the multipolarity of the excitation. In more detail, for the $^{18}$O projectile the excitation to the 2$_1$$^{+}$ state was considered in the coupling scheme using B(E$_2$; 0$^{+}$$\rightarrow$2$^{+}$)= 0.0043 e$^{2}$b$^{2}$ as reported in \cite{pritychenko}. For the $^{16}$O and $^{48}$Ti targets, inelastic excitations to the 3$_1$$^{-}$ and 2$_1$$^{+}$ states, respectively, were introduced in the calculation adopting the values of B(E$_3$; 0$^{+}$$\rightarrow$3$^{-}$)= 0.0015 e$^{2}$b$^{3}$ from Ref.\cite{kibedi} and B(E$_2$; 0$^{+}$$\rightarrow$2$^{+}$)= 0.072 e$^{2}$b$^{2}$ from Ref.\cite{raman}. The nuclear coupling potentials were derived following the same procedure as described in Ref. \cite{spp_med2}.\par
The spectroscopic amplitudes for the projectile and target overlaps were computed within the framework of shell-model \cite{shell1,shell2} using the code KSHELL \cite{kshell}. In particular, we have adopted two different effective Hamiltonians, namely the p-sd-mod \cite{psdmod1} and the SDPF-MU \cite{sdpfmu} interactions, both defined in a model space encompassing two major shells. The first one is a modified version of the PSDWBT interaction \cite{wbm} and is defined in a model space including the 1p and 2s-1d orbitals for both protons and neutrons with $^{4}$He as closed core. It has been already adopted in many of ours previous studies (e.g. \cite{cardozo,carbone2,linares_1n}) and is now used to compute the one-proton spectroscopic amplitudes involved in reactions with the $^{16}$O and $^{27}$Al targets. The second interaction, employed to calculate the spectroscopic amplitudes for the $\braket{^{47}Sc|^{48}Ti}$ overlaps, is constructed in a model space spanned by the proton and neutron 2s-1d and 1f-2p orbitals on top of the doubly magic $^{16}$O core. It is based on the V$_{Mu}$ potential \cite{vmu_pot} and has been widely adopted to study the effects of excitations across the sd-pf major shells as well as to compute the spectroscopic factors of nuclei in Ca region \cite{55ni,48ca,55sc,54ca}. However, in this case complete calculations are unfeasible due to the very large capacity of the adopted model space and, therefore, a truncation of the shell-model basis is required. In particular, we have considered only one particle-one hole cross-shell excitations, while all possible configurations within the 2s-1d and 1f-2p shells are taken into account. A list with the spectroscopic amplitudes, used in the reaction calculations, is presented in Tables I-III.
%
%
%
\begin{table}
\begin{center}
\caption{\textit{List of the one-proton spectroscopic amplitudes for the projectile overlaps used in the DWBA and CCBA calculations. The symbols \textbf{n}, \textbf{l} and \textbf{j} correspond to the principal quantum number, the orbital and the total angular momentum of the transferred proton orbitals, respectively.}}
\begin{tabular}{|c|c|c|c|}
\hline
Initial state & \parbox{1.3cm}{\centering{\textbf{nlj}}} & \parbox{2.5cm}{\centering{Final state}} & \parbox{2.0cm}{\centering{Spectroscopic amplitude}}\\
\hline
 & 2s$_{1/2}$ & $^{19}$F$_{g.s.}$ (1/2$^+$) & -0.554\\
 & 1p$_{1/2}$ & $^{19}$F$_{0.110}$ (1/2$^-$) & -0.244\\
 $^{18}$O$_{g.s.}$ (0$^+$) & 1d$_{5/2}$ & $^{19}$F$_{0.197}$ (5/2$^+$) & 0.664\\
 & 1p$_{3/2}$ & $^{19}$F$_{1.459}$ (3/2$^-$) & -0.011\\
 & 1d$_{3/2}$ & $^{19}$F$_{1.554}$ (3/2$^+$) & -0.424\\
 &	      &				    &	     \\
 & 1d$_{5/2}$ & $^{19}$F$_{g.s.}$ (1/2$^+$) & -0.586\\
 & 1d$_{3/2}$ & $^{19}$F$_{g.s.}$ (1/2$^+$) &  0.281\\
 & 1p$_{3/2}$ & $^{19}$F$_{0.110}$ (1/2$^-$) & 0.030\\
 & 1d$_{5/2}$ & $^{19}$F$_{0.197}$ (5/2$^+$) & 0.427\\
 & 1d$_{3/2}$ & $^{19}$F$_{0.197}$ (5/2$^+$) & -0.156\\
 $^{18}$O$_{1.982}$ (2$^+$) & 2s$_{1/2}$ & $^{19}$F$_{0.197}$ (5/2$^+$) & 0.311\\
 & 1p$_{3/2}$ & $^{19}$F$_{1.459}$ (3/2$^-$) & 0.002\\
 & 1p$_{1/2}$ & $^{19}$F$_{1.459}$ (3/2$^-$) & -0.164\\
 & 1d$_{5/2}$ & $^{19}$F$_{1.554}$ (3/2$^+$) & -0.315\\
 & 1d$_{3/2}$ & $^{19}$F$_{1.554}$ (3/2$^+$) & -0.319\\
 & 2s$_{1/2}$ & $^{19}$F$_{1.554}$ (3/2$^+$) & -0.354\\
\hline
\end{tabular}
\end{center}
\end{table}
%
%
%
\begin{table}
\begin{center}
\caption{\textit{List of the one-proton spectroscopic amplitudes for the target overlaps used in the DWBA and CCBA calculations for the reactions with the $^{16}$O and $^{27}$Al targets. The symbols \textbf{n}, \textbf{l} and \textbf{j} correspond to the principal quantum number, the orbital and the total angular momentum of the transferred proton orbitals, respectively.}}
\begin{tabular}{|c|c|c|c|}
\hline
Initial state & \parbox{1.3cm}{\centering{\textbf{nlj}}} & \parbox{2.0cm}{\centering{Final state}} & \parbox{2.0cm}{\centering{Spectroscopic amplitude}}\\
\hline
 & 1d$_{5/2}$ & $^{26}$Mg$_{g.s.}$ (0$^+$) & 0.525\\
 & 1d$_{5/2}$ & $^{26}$Mg$_{1.809}$ (2$^+$) & 0.915\\
 & 2s$_{1/2}$ & $^{26}$Mg$_{1.809}$ (2$^+$) & -0.140\\
 & 1d$_{3/2}$ & $^{26}$Mg$_{1.809}$ (2$^+$) & 0.028\\
 & 1d$_{5/2}$ & $^{26}$Mg$_{4.318}$ (4$^+$) & 1.069\\
 $^{27}$Al$_{g.s.}$ (5/2$^+$) & 1d$_{3/2}$ & $^{26}$Mg$_{4.318}$ (4$^+$) & 0.120\\
 & 1d$_{5/2}$ & $^{26}$Mg$_{4.332}$ (2$^+$) & 0.209\\
 & 1d$_{3/2}$ & $^{26}$Mg$_{4.332}$ (2$^+$) & 0.056\\
 & 2s$_{1/2}$ & $^{26}$Mg$_{4.332}$ (2$^+$) & 0.129\\
 & 1d$_{5/2}$ & $^{26}$Mg$_{4.350}$ (3$^+$) & 0.062\\
 & 1d$_{3/2}$ & $^{26}$Mg$_{4.350}$ (3$^+$) & 0.138\\
 & 2s$_{1/2}$ & $^{26}$Mg$_{4.350}$ (3$^+$) & 0.195\\
 &	      &				    &	     \\
 & 1p$_{1/2}$ & $^{15}$N$_{g.s.}$ (1/2$^-$) & -1.253\\
 $^{16}$O$_{g.s.}$ (0$^+$) & 1d$_{5/2}$ & $^{15}$N$_{5.270}$ (5/2$^+$) & 0.493\\
 & 2s$_{1/2}$ & $^{15}$N$_{5.299}$ (1/2$^+$) & 0.141\\
 & 1p$_{3/2}$ & $^{15}$N$_{6.324}$ (3/2$^-$) & 1.753\\
 &	      &				    &	     \\
 & 1d$_{5/2}$ & $^{15}$N$_{g.s.}$ (1/2$^-$) & 0.617\\
 & 1p$_{3/2}$ & $^{15}$N$_{5.270}$ (5/2$^+$) & 0.062\\
 $^{16}$O$_{6.130}$ (3$^-$) & 1p$_{1/2}$ & $^{15}$N$_{5.270}$ (5/2$^+$) & -0.834\\
 & 1d$_{5/2}$ & $^{15}$N$_{6.324}$ (3/2$^-$) & 0.160\\
 & 1d$_{3/2}$ & $^{15}$N$_{6.324}$ (3/2$^-$) & 0.268\\
\hline
\end{tabular}
\end{center}
\end{table}
%
%
%
\begin{table}
\begin{center}
\caption{\textit{List of the one-proton spectroscopic amplitudes for the target overlaps used in the DWBA and CCBA calculations for the $^{48}$Ti($^{18}$O,$^{19}$F)$^{47}$Sc reaction. The symbols \textbf{n}, \textbf{l} and \textbf{j} correspond to the principal quantum number, the orbital and the total angular momentum of the transferred proton orbitals, respectively.}}
\begin{tabular}{|c|c|c|c|}
\hline
Initial state & \parbox{1.3cm}{\centering{\textbf{nlj}}} & \parbox{2.0cm}{\centering{Final state}} & \parbox{2.0cm}{\centering{Spectroscopic amplitude}}\\
\hline
 & 1f$_{7/2}$ & $^{47}$Sc$_{g.s.}$ (7/2$^-$) & -1.325\\
 & 1d$_{3/2}$ & $^{47}$Sc$_{0.767}$ (3/2$^+$) & -1.536\\
 & 2p$_{3/2}$ & $^{47}$Sc$_{0.808}$ (3/2$^-$) & -0.306\\
 & 1f$_{5/2}$ & $^{47}$Sc$_{1.297}$ (5/2$^-$) & 0.043\\
 $^{48}$Ti$_{g.s.}$ (0$^+$) & 2s$_{1/2}$ & $^{47}$Sc$_{1.391}$ (1/2$^+$) & -1.094\\
 & 1d$_{5/2}$ & $^{47}$Sc$_{1.404}$ (5/2$^+$) & 0.122\\
 & 2s$_{1/2}$ & $^{47}$Sc$_{1.798}$ (1/2$^+$) & 0.332\\
 & 1d$_{3/2}$ & $^{47}$Sc$_{2.002}$ (3/2$^+$) & -0.039\\
 & 1d$_{5/2}$ & $^{47}$Sc$_{2.381}$ (5/2$^+$) & 0.422\\
 & 2s$_{1/2}$ & $^{47}$Sc$_{2.529}$ (1/2$^+$) & -0.050\\
 &	      &				    &	     \\
 & 1f$_{7/2}$ & $^{47}$Sc$_{g.s.}$ (7/2$^-$) & -0.793\\
 & 1f$_{5/2}$ & $^{47}$Sc$_{g.s.}$ (7/2$^-$) & -0.003\\
 & 2p$_{3/2}$ & $^{47}$Sc$_{g.s.}$ (7/2$^-$) & -0.253\\
 & 1d$_{3/2}$ & $^{47}$Sc$_{0.767}$ (3/2$^+$) & 0.377\\
 & 1d$_{5/2}$ & $^{47}$Sc$_{0.767}$ (3/2$^+$) & -0.083\\
 & 2s$_{1/2}$ & $^{47}$Sc$_{0.767}$ (3/2$^+$) & -0.247\\
 & 2p$_{3/2}$ & $^{47}$Sc$_{0.808}$ (3/2$^-$) & -0.113\\
 & 1f$_{7/2}$ & $^{47}$Sc$_{0.808}$ (3/2$^-$) & -0.612\\
 & 1f$_{5/2}$ & $^{47}$Sc$_{0.808}$ (3/2$^-$) & -0.025\\
 & 2p$_{1/2}$ & $^{47}$Sc$_{0.808}$ (3/2$^-$) & 0.004\\
& 1f$_{7/2}$ & $^{47}$Sc$_{1.297}$ (5/2$^-$) & -0.281\\
& 1f$_{5/2}$ & $^{47}$Sc$_{1.297}$ (5/2$^-$) & 0.008\\
& 2p$_{3/2}$ & $^{47}$Sc$_{1.297}$ (5/2$^-$) & -0.129\\
& 2p$_{1/2}$ & $^{47}$Sc$_{1.297}$ (5/2$^-$) & -0.014\\
 $^{48}$Ti$_{0.984}$ (2$^+$) & 1d$_{5/2}$ & $^{47}$Sc$_{1.391}$ (1/2$^+$) & 0.108\\
 & 1d$_{3/2}$ & $^{47}$Sc$_{1.391}$ (1/2$^+$) & 0.297\\
 & 1d$_{5/2}$ & $^{47}$Sc$_{1.404}$ (5/2$^+$) & -0.031\\
 & 1d$_{3/2}$ & $^{47}$Sc$_{1.404}$ (5/2$^+$) & -0.913\\
 & 2s$_{1/2}$ & $^{47}$Sc$_{1.404}$ (5/2$^+$) & 0.316\\
 & 1d$_{5/2}$ & $^{47}$Sc$_{1.798}$ (1/2$^+$) & -0.067\\
 & 1d$_{3/2}$ & $^{47}$Sc$_{1.798}$ (1/2$^+$) & 0.072\\
 & 1d$_{3/2}$ & $^{47}$Sc$_{2.002}$ (3/2$^+$) & -0.473\\
 & 1d$_{5/2}$ & $^{47}$Sc$_{2.002}$ (3/2$^+$) & 0.114\\
 & 2s$_{1/2}$ & $^{47}$Sc$_{2.002}$ (3/2$^+$) & -0.600\\
 & 1d$_{5/2}$ & $^{47}$Sc$_{2.381}$ (5/2$^+$) & 0.072\\
 & 1d$_{3/2}$ & $^{47}$Sc$_{2.381}$ (5/2$^+$) & -0.248\\
 & 2s$_{1/2}$ & $^{47}$Sc$_{2.381}$ (5/2$^+$) & -0.786\\
 & 1d$_{5/2}$ & $^{47}$Sc$_{2.529}$ (1/2$^+$) & -0.031\\
 & 1d$_{3/2}$ & $^{47}$Sc$_{2.529}$ (1/2$^+$) & -0.014\\
\hline
\end{tabular}
\end{center}
\end{table}
\section{Results and discussion}
Considering the case of the $^{16}$O($^{18}$O,$^{19}$F)$^{15}$N reaction, as it was stated above, the excitation energy spectrum up to $\approx$ 5 MeV underlines the fingerprint of the $^{19}$F nucleus. The ground state of $^{15}$N is mainly described as a 1p$_{1/2}$ hole coupled to the $^{16}$O ground state and thus, the $^{16}$O$_{g.s.}$ $\rightarrow$ $^{15}$N$_{g.s.}$ transition proceeds via a proton removal from the 1p$_{1/2}$ shell \cite{nitro_states,nitro_states2,nitro_states3}. The 1p$_{1/2}$ shell of $^{16}$O is fully occupied and the one-proton pickup proceeds via the removal of one of the two indistinguishable valence protons. Thus, it is expected that the absolute value of the spectroscopic amplitude for the $\braket{^{15}N_{g.s.}|^{16}O_{g.s.}}$ overlap should be closed to $\sqrt 2$. Indeed, the absolute value for the spectroscopic amplitude predicted by our shell-model calculation is equal to 1.2532 in agreement with those reported in Ref.\cite{nitro_states3} from (d,$^{3}$He) experiments. As a result, a comparison between experimental and theoretical angular distribution data corresponding to the first two peaks of Fig.5, provides the ground for testing the accuracy of the calculated spectroscopic amplitudes for the projectile overlaps. To this direction, DWBA and CCBA calculations for the one-proton transfer reaction populating $^{15}$N in its ground state and the low-lying states of $^{19}$F(see Table I) were performed and the results are presented in Fig.6. In all cases, the agreement between experimental and theoretical angular distribution data is very good. The inclusion of inelastic excitations of projectile and target in the coupling scheme improves in most of the cases the agreement with the experimental data but the case of the ground state region where the CCBA prediction slightly overestimates the experimental cross-sections. The striking example of the importance of projectile and target excitations is demonstrated in Fig.6b. Although both DWBA and CCBA calculations are in quantitative agreement with the experimental data, it is evident that only the latter is able to reproduce the oscillatory pattern of the angular distribution. This result points also to a small, yet significant, contribution from core excitation configurations in the population of the \(\frac{3}{2}\)$^{-}$ and \(\frac{3}{2}\)$^{+}$ states of $^{19}$F. At this point, it should be mentioned that additional calculations by introducing the \(\frac{5}{2}\)$^{-}$ 1.346 MeV state of $^{19}$F in the coupling scheme were performed. However, due to the small values in the spectroscopic amplitudes for this transition the theoretical cross-sections presented deviations of about 1\%. Therefore, this state was no longer considered in the calculations. In general, both calculations reproduce adequately-well the experimental data without need for any additional normalization factor \cite{dwba_sf3,ccba_effect2,unhappy,unhappy3,unhappy4}. The good agreement between experimental and theoretical cross-section gives further support to the validity of the calculated spectroscopic amplitudes as well as the adopted optical potentials for the description of the initial and final-state interactions.\par
Once the spectroscopic amplitudes for the $\braket{^{19}F|^{18}O}$ overlaps were established, similar calculations for the $^{27}$Al($^{18}$O,$^{19}$F)$^{26}$Mg and $^{48}$Ti($^{18}$O,$^{19}$F)$^{47}$Sc reactions were performed and the results are compared to the experimental data in Figs. 3 and 8, respectively. Starting from the case of $^{27}$Al, the theoretical angular distributions are in very good agreement with the experimental data but with a tendency to slightly underestimate the magnitude of the experimental cross-sections at the high excited states. The coupling to the \(2\)$_{1}$$^{+}$ state of $^{18}$O is found to be weak but its inclusion improves the description of the angular distribution data. The shape of the measured angular distribution is rather structure-less reflecting the strong contribution from the 1d$_{5/2}$ transfer to the \(\frac{5}{2}\)$^{+}$ state of $^{19}$F. This is consistent with the findings from ($\alpha$,t) experiments \cite{states_yasue_f19}. The theoretical calculation for the $^{48}$Ti($^{18}$O,$^{19}$F)$^{47}$Sc reaction was more demanding, since a large number of nuclear states of $^{47}$Sc was introduced, thus presenting a good test for adopted spectroscopic amplitudes. Despite that, once more the predicted angular distribution compares very well with the experimental data, as seen in Fig.8. \par
Having completed the analysis for the three reactions under study, it is worthwhile to notice that the leading role in the projectile transitions for the excitation energy range under investigation is always provided by the transition to the \(\frac{5}{2}\)$^{+}$ state at 0.197 MeV of $^{19}$F, regardless the transition undertaken from the target. 
\section{Summary and conclusions}
Cross-section measurements for the one-proton pickup reactions induced by $^{18}$O on $^{16}$O, $^{27}$Al and $^{48}$Ti targets were performed at INFN-LNS. The $^{19}$F ejectiles were detected by the MAGNEX magnetic spectrometer spanning an angular range between \ang{3} and \ang{15} in the laboratory reference frame. It has been demonstrated that adopting the DWBA and/or CCBA formalism we are able to describe very accurately the one-proton transfer angular distribution data, without the need for any re-normalization factor of the theoretical curves. The couplings to the low-lying states of the projectile and target nuclei were found to be weak but in most of the cases improve the description of the experimental data. It should be pointed out that these couplings were not appropriately constrained, since elastic and inelastic scattering data were not available, but is in our intentions to perform elastic and inelastic scattering measurements for the same system in the near future. Considering that the transition amplitudes in the DWBA approach are strongly dependent on the spectroscopic factors, being also rather sensitive to the spatial distribution of the projectile-target interaction, the good agreement observed between the experimental and theoretical cross-sections supports the validity of the adopted optical potentials as well as the accuracy of the spectroscopic amplitudes which were derived from shell-model calculations. This result is very important for the NUMEN project. In the quantum mechanical description of the DCE mechanism the wave functions which enter in the expression of the transition amplitude are the distorted waves at the entrance and exit channels and the transition densities. Like in the case of transfer reactions, these distorted waves are the solution of the Schroedinger equation using the optical potentials, while the calculation of the transition densities relies on nuclear structure models. Therefore, transfer reactions can provide the testing ground for the reaction and structure models, before moving on the description of more complicated reaction mechanisms like the SCE and DCE ones. Finally, the results of the present analysis together with those from the ($^{18}$O,$^{17}$O) one in the $^{18}$O + $^{48}$Ti reaction will clarify the degree of competition between the direct SCE mechanism ($^{18}$O,$^{18}$F) and the sequential nucleon transfer, with the former being particularly important in order to constrain the DCE sequential mechanism in the $^{18}$O$\rightarrow$$^{18}$F$\rightarrow$$^{18}$Ne transition. 
\section*{Acknowledgments}
We warmly acknowledge the operators of INFN-LNS Accelerator for the production and delivery of the $^{18}$O beam and their support throughout the experiment. The research leading to these results was partially funded by the European Research Council (ERC) under the European Union’s Horizon 2020 Research and Innovation Programme (Grant Agreement No 714625). We also acknowledge the CINECA award under the ISCRA initiative (code HP10B51E4M) and through the INFN-CINECA agreement for the availability of high performance computing resources and support. One of us (G.D.G.) acknowledges the support by the funding program VALERE of Univerit\`a degli Studi della Campania "Luigi Vanvitelli". One of us (S.B.) acknowledges support from the Alexander von Humboldt foundation. We also acknowledge partial financial support from CNPq, FAPERJ, FAPESP (proc. no. 2019/07767-1) and Instituto Nacional de Ci\^{e}ncia e Tecnologia - F\'isica Nuclear e Aplica\c{c}\~{o}es (INCT-FNA, proc. no. 464898/2014-5), Brazil.
\end{document}